\documentclass[twocolumn,showpacs,preprintnumbers,amsmath,amssymb]{revtex4}

\usepackage{graphicx}
\usepackage{dcolumn}
\usepackage{bm}

\begin{document}

\title{Granular Flows in Split-Bottom Geometries}

\author{Joshua A. Dijksman}

 \affiliation{Kamerlingh Onnes Lab, Universiteit Leiden, Postbus
9504, 2300 RA Leiden, The Netherlands}

\author{Martin van Hecke}

\affiliation{Kamerlingh Onnes Lab, Universiteit Leiden, Postbus
9504, 2300 RA Leiden, The Netherlands}

\date{\today}

\begin{abstract}
There is a simple and general experimental protocol to generate
slow granular flows that exhibit wide shear zones, qualitatively
different from the narrow shear bands that are usually observed in
granular materials . The essence is to drive the granular medium
not from the sidewalls, but to split the bottom of the container
that supports the grains in two parts and slide these parts past
each other. Here we review the main features of granular flows in
such split-bottom geometries.
\end{abstract}

 \maketitle

Granular media exhibit a complex mixture of solid and fluid-like
behavior, often hard to predict or capture in models. Perhaps the
most striking feature of granular flows is their tendency to
localize in narrow shear bands~\cite{2010_annurevfluid_schall}. A
decent model of grain flows should be able to capture, and
preferably, predict this type of behavior, but at present there is
no {\em general} approach which, for given geometry, driving
strength and grain properties, predicts the ensuing flow fields.

In recent years, much progress has been made for fast flows, such
as avalanche flows down an incline, where large flowing zones
form. Microscopically, momentum exchange then takes place by a
mixture of collisions and enduring contacts. This allows the
definition of a dimensionless parameter $I$, the inertial number,
which characterizes the local ''rapidity'' of the flow. A local
relation between stresses, strain-rates and $I$ then successfully
captures many aspects of these rapid granular flows
\cite{2004_epje_gdrmidi,
2006_nature_jop,2008_annurevfluid_forterre}.

In contrast, the situation for slow flows, such as those made by
slowly shearing the boundaries of a container containing grains,
is still wide open. The averaged stresses and flow profiles become
essentially independent of the flow rate, so that constitutive
relations based on relating stresses and strain rates are unlikely
to capture the full physics. In this regime, shear banding is
generally very strong, with shear bands having a typical thickness
of five to ten grain diameters. These shear bands often localize
near the moving boundary. For a recent review,
see~\cite{2010_annurevfluid_schall}. Experimental handles for
probing this shear localization appears to be limited, since shear
banding appears so robust. For example, granular flows in Couette
cells always show the formation of a narrow shear band near the
inner cylinder, irrespective of dimensionality, driving rate, or
details of the geometry~\cite{2000_nature_mueth}.

In this regime of slow flows, the inertial number $I$ tends to
zero and momentum transfer is dominated by enduring contacts. Soil
mechanics is then a natural starting point to describe these
flows, and both rate independence and shear banding are consistent
with a Mohr-Coulomb picture where the friction laws acting at the
grain scale are translated to the stresses acting at
coarse-grained level. The idea is that when the ratio of shear to
normal stresses is below the yielding threshold, grains remain
quiescent, while in slowly flowing regions the shear stresses will
be given by a (lower) dynamical yield stress. This way of thinking
readily captures the maximal slope of dry sand piles. However, the
steep gradients associated with narrow shear bands are difficult
to capture by a continuum theory, and shear bands often are
described as having zero width \cite{2004_prl_unger}.

Shearbands, then, are not always narrow. In this paper we will
review recent experiments, numerical work and theoretical
descriptions of wide shear zones which have been generated in
so-called split-bottom geometries. The essence is to drive the
granular medium not from the sidewalls, but to split the bottom of
the container that supports the grains in two parts that slide
past each other. By taking advantage of gravity to drive the
granulate from the sliding discontinuity in the bottom support of
the grain layer, one effectively pins a wide shear zone away from
the sidewalls. The resulting grain flows are smooth and robust,
with both velocity profiles and the location of the shear zones
exhibiting simple, grain independent properties.

The outline of this paper is as follows. In
section~\ref{sec:intro:splibo} we will review the results of
recent experiments and numerical work on the flows which have been
generated in these special flow geometries. In
section~\ref{sec:intro:theory} we will discuss the theoretical
ideas proposed to capture the flows observed in these geometries.

\section{Slow Flows in the Split-Bottom Geometry: Phenomenology}\label{sec:intro:splibo}
        \begin{figure}[tb!]
            \begin{center}
                \includegraphics[width=8cm]{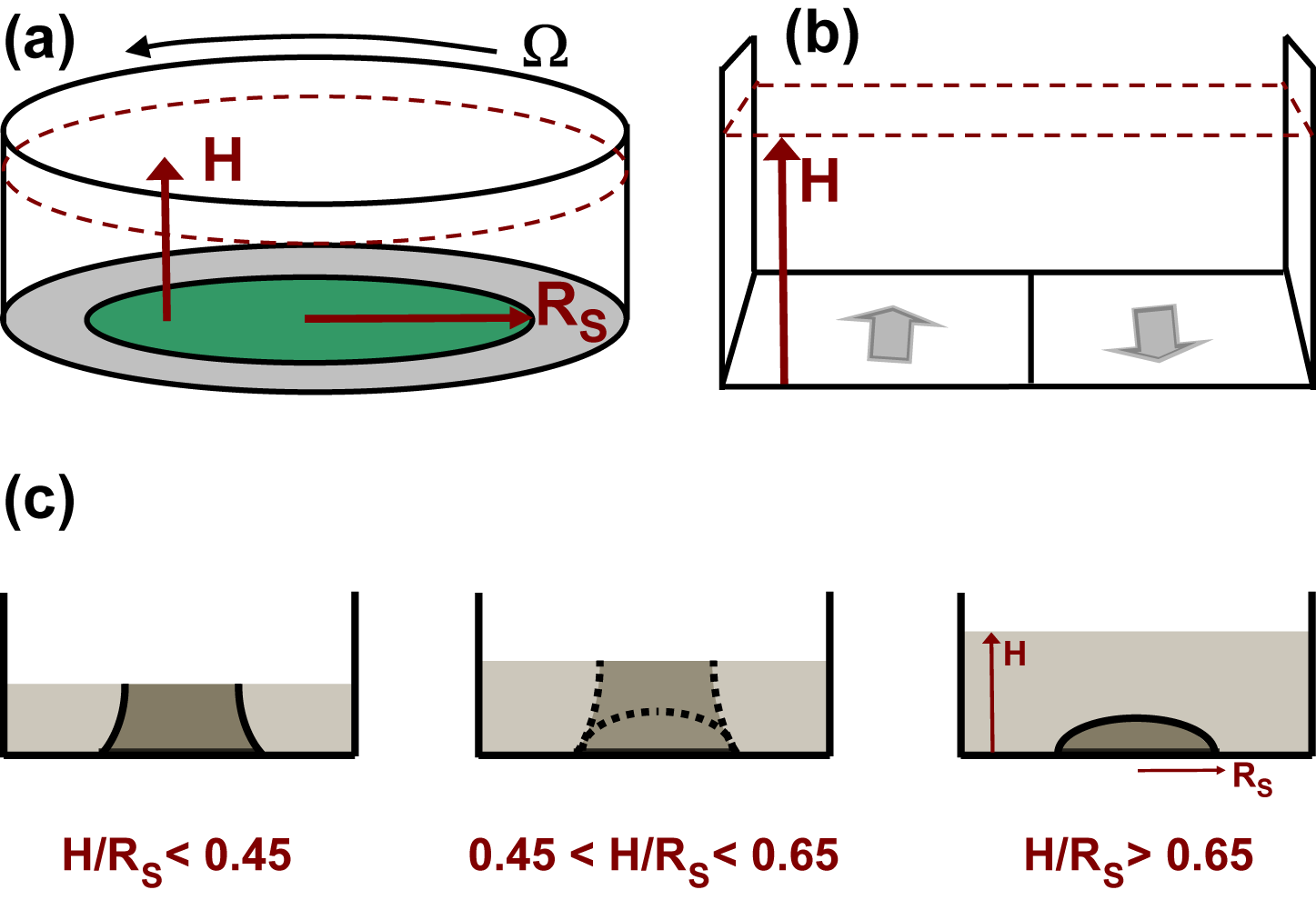}
                \caption{(a) Cylindrical
                split-bottom geometry, showing a disc of radius $R_s$  at the
                bottom of a granular layer of depth $H$. Here, the outer cylinder
                rotates with rate  $\Omega$ and the bottom disc is kept fixed --
                A similar geometry, with fixed outer cylinder and rotating
                disc is also frequently encountered. (b) Linear split-bottom
                geometry, where a container is split along a straight line in its
                bottom. This geometry can be seen as the $R_s \rightarrow \infty$
                limit of the cylindrical
                cell. (c) The transition in flow structure from shallow to deep flows in
                the cylindrical split-bottom geometry. In the dark grey
                region the material essentially co-moves with the
                disk.}\label{fig:intro:splibogeom}
            \end{center}
        \end{figure}

        \subsection{General Description}
        In this section, we focus on the rate independent regime which is reached
        for slow enough driving. Two variants
        of the split-bottom geometry will be encountered: in experiments
        one typically employs a cylindrical split-bottom shear cell,
        consisting of a bucket, at the bottom of which a disc rotates with
        respect to the bucket (Fig.~\ref{fig:intro:splibogeom}a)
        \cite{2003_nature_fenistein,2004_prl_fenistein,2006_prl_fenistein,2006_prl_cheng}, while for
        theoretical studies the linear split-bottom cell
        with periodic boundary conditions is more
        convenient (Fig.~\ref{fig:intro:splibogeom}b) \cite{2006_pre_depken,2007_epl_depken,2007_pre_ries,2008_pre_jagla}.

        \subsection{Parameters and Regimes}
        The cylindrical split-bottom geometry is characterized by three
        parameters. The radius of the bottom disc $R_s$ and its rotation
        rate $\Omega$ are generally fixed in a set of experiments, and the
        relative motion of disc with respect to the cylindrical container
        drives the flow. The thickness of the granular layer, $H$, is the
        control parameter that typically is scanned in a series of
        experiments. Note that the radius of the container appears
        immaterial, as long as it is sufficiently large; 25\% larger than
        $R_s$ appears to be sufficient~\cite{2004_prl_fenistein}.

        We denote the ratio of the averaged azimuthal velocity of the grains
        $v_{\theta}(r)/r$ and the disk rotation speed $\Omega$ by $\omega$;
        $\omega=0$ thus corresponding to stationary
        grains, while $\omega=1$ corresponds to grains co-moving with the
        driving. For the small $\Omega$ of interest here (typically less
        than 0.1 s$^{-1}$), the flow profiles $\omega(r,z)$ are
        independent of $\Omega$
        -- the flow is rate independent, and transients are short lived.
        Since centrifugal forces are negligible for typical sizes of $R_s$
        (typically a few cm), one may also fix the disc and rotate the
        bucket, and essentially obtain the same sort of flows with
        $\tilde{\omega}(r,z)=1-\omega(r,z)$
        -- see \cite{2003_nature_fenistein,2004_prl_fenistein,2006_prl_fenistein}.
        The two parameters $H$ and $R_s$ set the large-scale
        structure of the flow.

        When the disk rotates, a shear zone propagates from the slip
        position $R_s$ upwards and inwards. The qualitative flow behavior
        is governed by the ratio $H/R_s$, and three regimes can be
        distinguished -- Fig.~\ref{fig:intro:splibogeom}c.
        A regime of shallow layers is found for $H/R_s <
        0.45$, and here the shear zone reaches the free surface. The
        three-dimensional shape of the shear zones is roughly that of the cone
        of a trumpet, with the front of the trumpet buried upside down in
        the sand. Another regime of deep layers plays a role for $H/R_s
        > 0.65$, and here the shear zone essentially forms a dome-like
        structure in the bulk of the material; little or no shear is
        observed at the free surface. In between there is an intermediate
        regime, where the shear in the bulk of the material is a mix
        between the trumpet and dome-like shape.

        \begin{figure}[tb!]
            \begin{center}
                \includegraphics[width = 8cm]{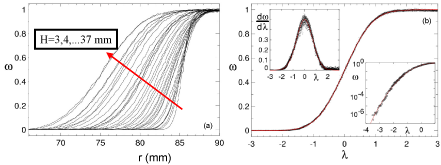}
                \caption{(a) Surface
                flows for glass beads of diameter 300 $\mu$m and a range of
                filling heights $H$ as indicated are well described by an error
                function (fit not shown) - $R_s=85$ mm here, and the outer
                cylinder is rotating. (b) Collapse of the surface flow profiles
                shown in (a) and comparison to error function. The rescaled radial
                coordinate $\lambda$ is defined as $(r-R_c)/W$. Top inset: strain
                rates are Gaussian. Bottom inset: the tail of flow profile
                corresponds well to the Gaussian tail of the error
                function. Figure adapted from Ref.~\cite{2003_nature_fenistein}.}\label{fig:intro:erf}
            \end{center}
        \end{figure}

        \subsection{Surface Flow}\label{subsec:intro:surfaceflow}

        \textbf{Shallow layers --- }
        We first focus on the flow observed at the free surface. For
        shallow layers, a narrow shear zone develops above the split at
        $R_s$, and when $H$ is increased, the shear zone observed at the
        surface broadens continuously and without any apparent bound.
        Additionally, with increasing $H$, the shear zone shifts away from
        $R_s$ towards the center of the shear cell (Fig.~2a).

        After proper rescaling, all bulk profiles collapse on a universal
        curve which is extremely well fitted by an error function:

        \begin{equation}
            \omega(r)=\frac{1}{2}+\frac{1}{2}erf\{\frac{r-R_c}{W}\}, \label{eq:intro:erf}
        \end{equation}

        where $erf$ denotes the error function, $r$ is the radial
        coordinate, $R_c$ the center of the shearband (where
        $\omega(r)=0.5$) and $W$ the width of the shearband (Fig.~\ref{fig:intro:erf}).
        Accurate measurement of the tail of the velocity profile further
        validate Eq. (\ref{eq:intro:erf}), and rule out an exponential tail of the
        velocity profile (Fig.~\ref{fig:intro:erf}b). The strain rate is therefore Gaussian,
        and the shear zones are completely determined by their centers
        $R_c$ and widths $W$.

        Particle shape does not influence the functional form of the
        velocity profiles~\cite{2004_prl_fenistein}, in contrast to the particle dependence found
        for wall-localized shear bands in a Couette cell \cite{2000_nature_mueth}. For
        these, the vicinity of the wall induces
        particle layering, in particular for monodisperse mixtures.
        Apparently such layering effects play
        no role for these bulk shear zones, where it should be noted that the effects of
        shear-induced ordering of particles with larger aspect ratios
        has not yet been investigated.

        \begin{figure}[tb!]
            \begin{center}
                \includegraphics[width=8cm]{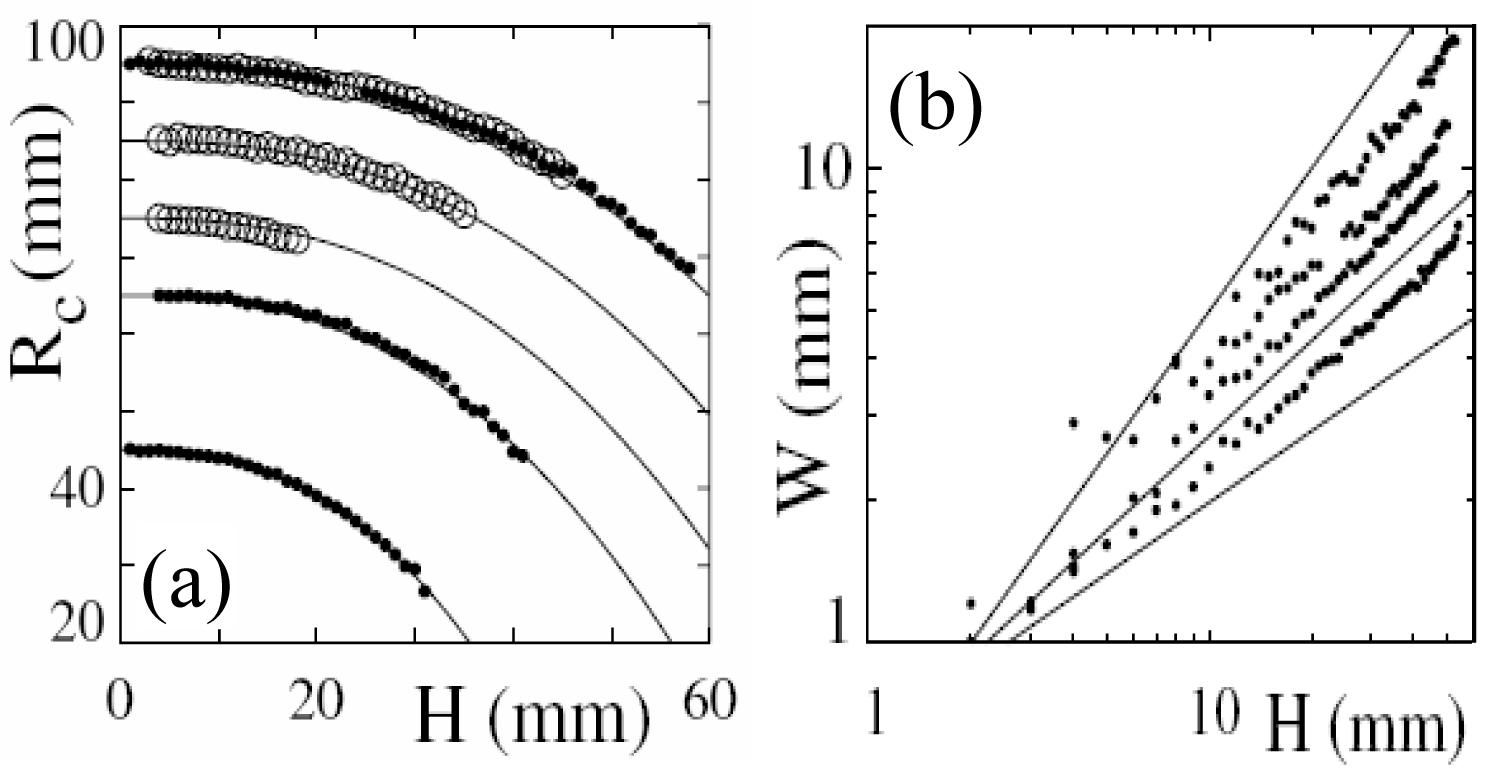}
                \caption{(a) Shear zone positions $R_c$ versus $H$,
                where $R_s$ = 95, 85, 75, 65, and 45 mm. Lines correspond to
                Eq.~\ref{eq:intro:rceq}. (b) Log-log plots of $W$ for spherical glass beads
                of increasing sizes (ranging from average diameter 300 $\mu$m to 2
                mm) for $R_s = 95$ mm. The lines shown in (b) correspond to exponents
                of 1/2, 2/3, 1. Figure adapted from \cite{2004_prl_fenistein}.}
                \label{fig:intro:shearzonepos}
            \end{center}
        \end{figure}

        Remarkably, the center of the shear zone, $R_c$, turns out to be
        independent of the material used (Fig.~\ref{fig:intro:shearzonepos}a). Therefore, the only
        relevant length-scales for $R_c$ appear to be $H$ and $R_s$. We
        find that the dimensionless 'displacement' of the shear zone,
        $(R_s-R_c)/R_s$, is a function of the dimensionless height
        ($H/R_s$) only. The simple relation
        \begin{equation}
            (R_s-R_c)/R_s = (H/R_s)^{5/2}\label{eq:intro:rceq}
        \end{equation}
        fits the data well (Fig.~\ref{fig:intro:shearzonepos}a).

        The relevant length scale for the shear zone width $W$ defined above is given by the grain
        properties, and is independent of $R_s$ (Fig.~\ref{fig:intro:shearzonepos}b).
        Grains shape, size,
        and type also influence $W(H)$: irregular particles display
        smaller shear zones than spherical ones of similar diameter. The
        best available experimental data shows that
        \begin{equation}
            W/d \sim (H/d)^{2/3}~, \label{eq:intro:weq}
        \end{equation}
        where $d$ denotes the particle diameter. Although this scaling has
        not been checked over more than a decade, the exponent is clearly not equal
        to $1/2$ or one. As yet there is no explanation for this scaling.

        The available numerical data coming from molecular dynamics
        simulations essentially confirm this picture
        \cite{2006_prl_cheng,2007_epl_depken,2007_pre_ries,2008_partscitech_luding}: the surface flows in
        split-bottom geometries for $H/R_s<0.45$ are given by
        Eqs.~\ref{eq:intro:erf}-\ref{eq:intro:weq}. Only the absolute width of the shear
        zone at the surface $W(z=H)$ remains as a fit parameter, but once this width has been
        measured for a single value of $H/R_s$, Eq.~\ref{eq:intro:weq} can be used
        to estimate the width for the whole range of
        $H/R_s<0.45$.\\

        \textbf{Deep Layers --- }
        When $H/R_s$ is small, the core material rests on and co-moves with the
        center disc. With increasing $H/R_s$, the width of the shear zone
        grows continuously, and its location moves inward towards the
        central region (Eq.~\ref{eq:intro:rceq}). This implies that for deep layers
        qualitatively different flow patterns can be expected to occur.

        \begin{figure}[tb!]
            \begin{center}
                \includegraphics[width=8cm]{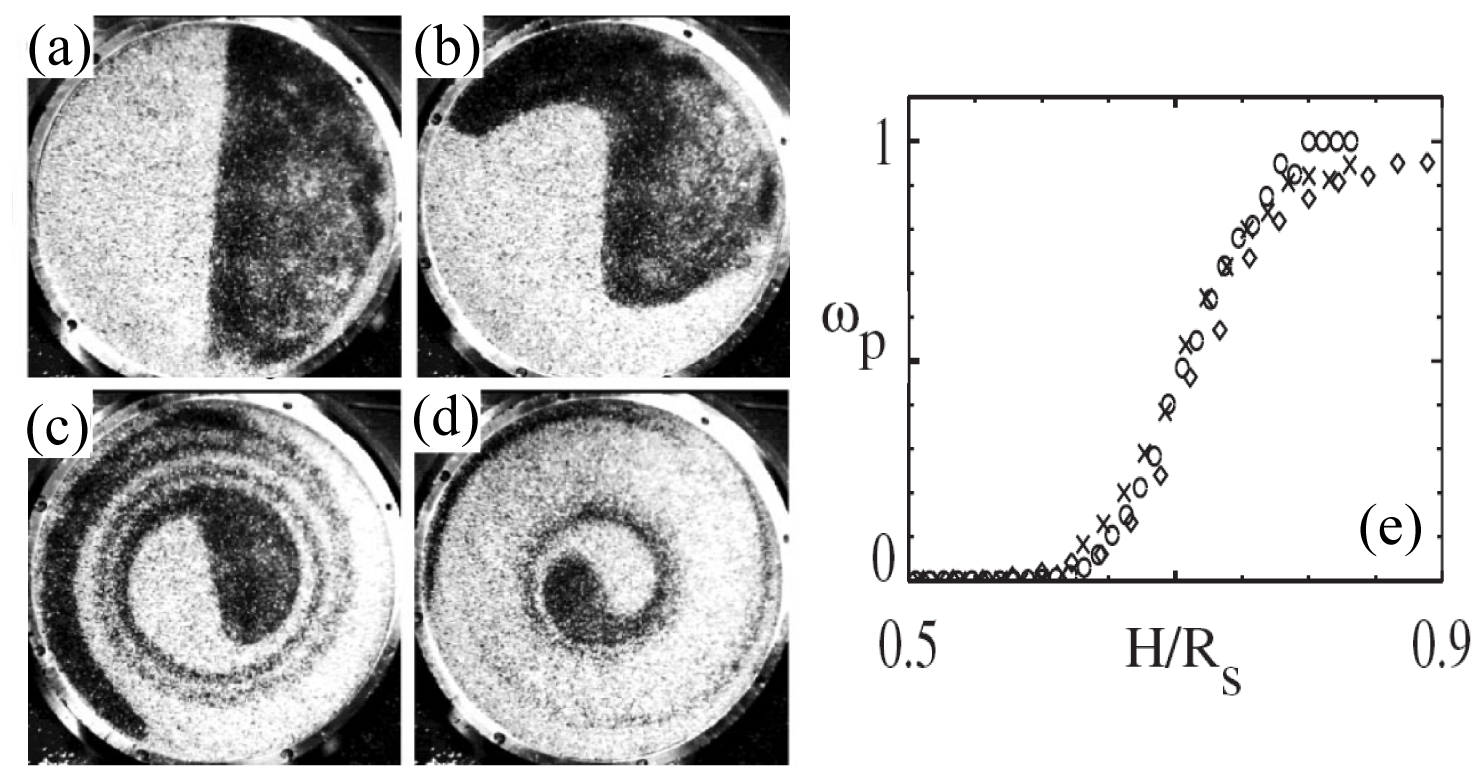}
                \caption{Core precession
                in a split-bottomed geometry. (a-d) Series of snapshots of top
                views of a setup with stationary disc and rotating outer cylinder
                (for $R_s = 95$ mm, $H = 60$ mm, and rotation rate $\Omega=
                0.024$ rps), where colored particles sprinkled on the surface
                illustrate the core precession for t = 0 s (a), t = 10 s (b), t =
                100 s (c) and t = 1000 s (d). (e) Data collapse of the precession
                rate $\omega_p$ for $R_s = 45$ mm (diamonds), $R_s$ = 65 mm (x)
                and $R_s = 95$ mm (circles) when plotted as a function of $H/R_s$.
                Figure adapted from Ref.~\cite{2006_prl_fenistein}.}\label{fig:intro:coreprec}
            \end{center}
        \end{figure}

        The most striking feature of these flows is that the core, as
        observed at the free surface,  precesses with respect to the
        bottom disc for $H/R_s \gtrsim 0.65$ --- hence material in the central part of the system no
        longer rests on the disc, and there is \emph{torsional failure} of the
        core. Precession is not simply a consequence of the overlap of
        two opposing shear zones, since before being eroded by shear, the
        inner core rotates as a solid blob for an appreciable time
        (Fig.~\ref{fig:intro:coreprec}a-d).

        The precession rate $\omega_p$ is defined as  the limit of
        $\omega(r)$ for $r$ going to zero, where we assume, for
        simplicity, that the outer bucket rotates with rate $\Omega$ and
        the bottom disc is kept fixed, as in \cite{2006_prl_fenistein} --
        consistent results are found in a setup where the disc was rotated
        and the outer cylinder kept fixed \cite{2006_prl_cheng}. For various slip
        radii, the onset height for precession grows with $R_s$, and the
        data for $\omega_p$ collapses when plotted as a function of $H/Rs$,
        see Fig.~\ref{fig:intro:coreprec}e). When $H/R_s$ becomes of order
        one, the whole surface
        rotates rigidly with the rotating drum and all shear takes place
        in the bulk of the material; on the other hand, for $H/R_s<0.65$,
        hardly any precession can be observed.\\

        \begin{figure}[tb!]
            \begin{center}
                \includegraphics[width=8cm]{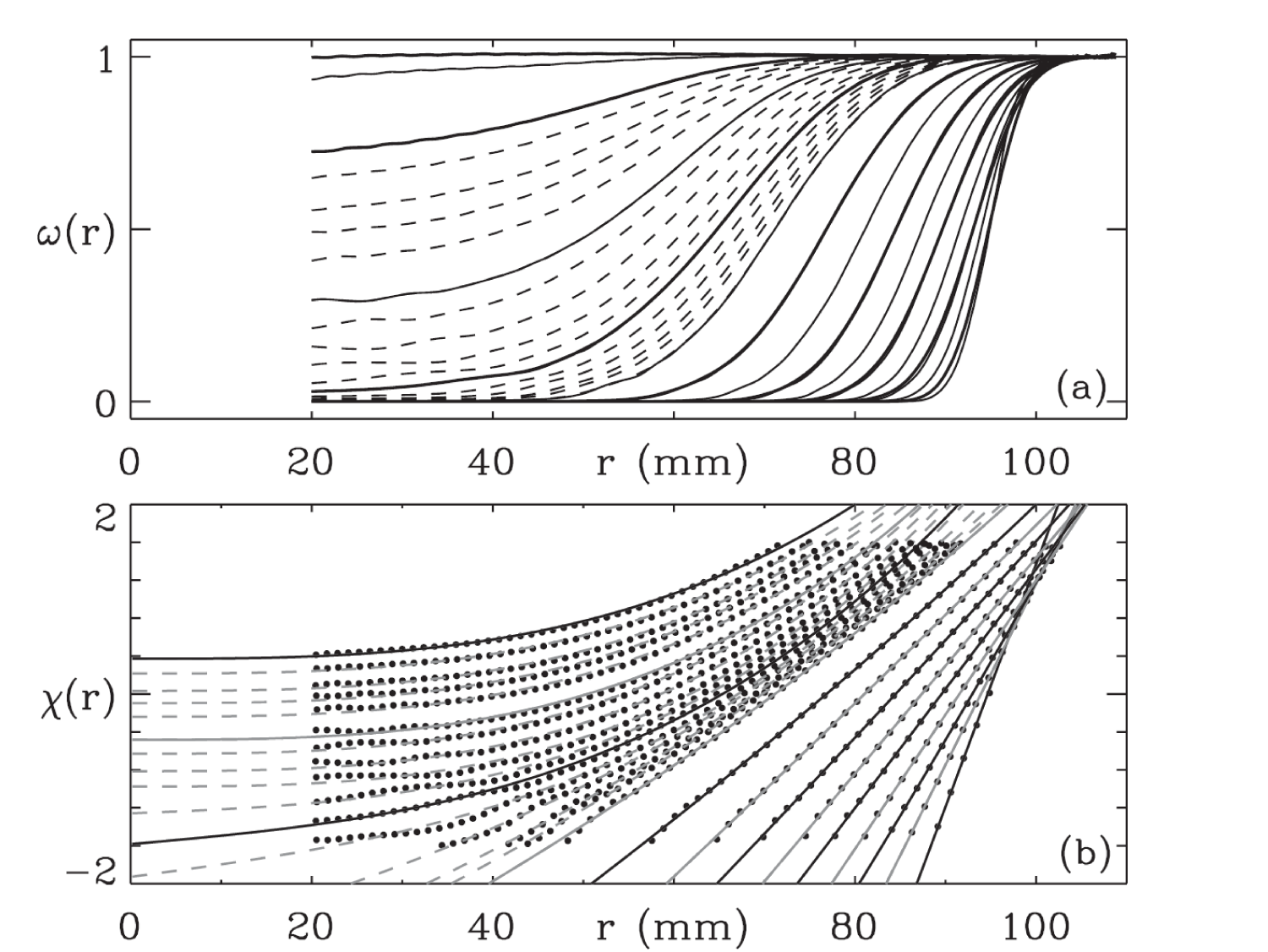}
                \caption{Surface velocity
                profiles $\omega(r)$ for $R_s= 95$ mm and increasing layer depth
                $H$. Thick curves: $H = 10, 20 \dots  80$ mm; Thin curves $H = 15,
                25,\dots  75$ mm; Dashed curves $H = 56, 57, \dots  69$ mm. (a)
                Precession gradually sets in for $H > 60$ mm. (b) Corresponding
                profiles of $\chi(r)$ (dots, see Eq.~\ref{eq:intro:nerf}), compared to cubic fits given by
                Eq.~\ref{eq:intro:cubfit} (curves). Figure adapted from~\cite{2006_prl_fenistein}.}
            \end{center}
        \end{figure}

        \textbf{Intermediate regime ---}
        In the intermediate regime, $0.45<H/R_s<0.65$, a precursor to the
        transition to precession can be observed, since a careful analysis
        reveals that the surface velocity profiles $\omega(r)$
        increasingly become asymmetric for $H/r_s>0.45$. In fact, one can
        generalize Eq.~\ref{eq:intro:erf} by writing
        \begin{equation}
            \omega(r) = \frac{1}{2}+\frac{1}{2}erf(\chi(r))~, \label{eq:intro:nerf}
        \end{equation}
        and by fitting the data for $\omega(r)$ over the whole range of
        $H/R_s$ to this equation (Fig.~5), one finds that $\chi(r)$ can be
        fitted well by a cubic polynomial of the form
        \begin{equation}
            \chi(r)=a_0+a_1 ~r +a_3 ~r^3~.\label{eq:intro:cubfit}
        \end{equation}
        For shallow layers, $a_3=0$, and $a_0$ and $a_1$ follow from the
        scaling laws Eqs.~\ref{eq:intro:rceq} and \ref{eq:intro:weq}. For $0.45<H/R_s<0.65$,
        $a_3$ starts to grow and governs the symmetry breaking of the flow
        profiles, while for deep layers ($H/R_s>0.65$), $a_1$ tends to
        zero, and a two parameter fit describes the flow profiles well
        again \cite{2006_prl_fenistein}.

        \subsection{Bulk Flow}\label{subsec:intro:bulk}

        \textbf{Shallow flows -- }
        The bulk structure of
        granular flows is harder to access, but by now, we have
        information on split-bottom flows from experiments that bury and
        excavate colored beads~\cite{2004_prl_fenistein}, Magnetic Resonance Imaging
        (MRI)~\cite{2006_prl_cheng, 2008_epl_sakaie} and numerical
        simulations~\cite{2006_prl_cheng,2007_epl_depken,2007_pre_ries,2008_partscitech_luding} (see Fig.~6).
        First, for shallow
        layers, the flow profiles at fixed depth $z$ below the surface $H$
        still takes an error function form, which allows us to characterize
        $\omega(r)$ at fixed $z$ with the same two parameters $R_c$ and
        $W$ as before.

        The location of the shear zones in the bulk where
        found to be consistent with a scaling argument put forward by
        Unger {\em et al.}. The idea is follows: Eq.~\ref{eq:intro:rceq} gives the
        location $R_c$ at the free surface. Then to obtain $R_c$ at depth
        $z$, one imagines a systems with a depth of $H-z$, and by
        inverting Eq.~\ref{eq:intro:rceq}, obtains where the split would have to be
        in a system of depth of $H-z$ for the surface location to be as
        given \cite{2004_prl_unger}. Identifying the split at depth $H-z$ with
        the center of the shearband $R_s(H-z)$, this yields:

        \begin{equation}
            z=H-R_c\left[1-R_c/R_s(1-H/R_s)^{2.5})\right]^{1/2.5}~.\label{eq:intro:rc}
        \end{equation}

        \noindent The width $W(z)$ of the shear zones is harder to obtain reliably,
        but the best available experimental data suggest a power law
        growth of the form $W\sim z^\alpha$, where $\alpha$ is less than
        1/2 and more than 1/4 \cite{2004_prl_fenistein,2006_pre_depken}. More recent
        numerical studies \cite{2007_pre_ries} found that $W(z)$ can also be well
        described by a ``quarter circle'' curve of the form
        \begin{equation}
            W(z)=W(z=H)\sqrt{1-(1-z/H)^2}~.\label{eq:intro:width}
        \end{equation}

\begin{figure}[tb!]
            \begin{center}
                \includegraphics[width=8cm]{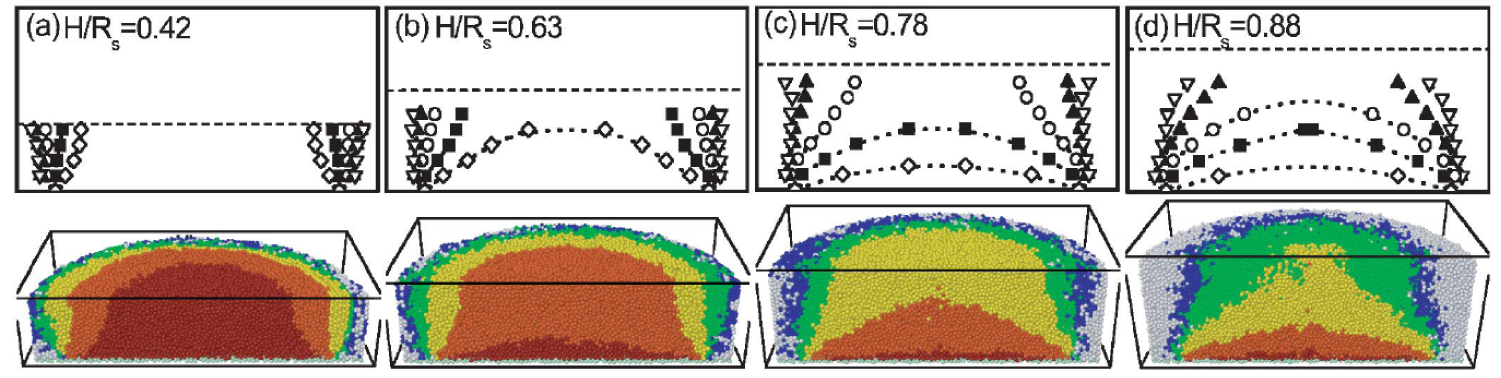}
                \caption{Contours of
                constant angular velocity $\omega$, for different filling height
                H. Upper panels: MRI experiments: $\omega= 0.84$ (diamonds), 0.24
                (squares), $2.4 \times 10^{-2}$  (circles), $2.4 \times 10^{-3}$
                (triangles), and 2.4 $\times 10^{-4}$ (inversed triangles). Dashed
                lines indicate H and dotted lines are guides to the eye. Lower
                panels: simulations. Color is used to identify velocity ranges:
                dark red, $\omega \in [0.84,1]$; orange, $\omega \in [0.24,
                0.84]$; yellow, $\omega \in [2.4\times 10^{-2}, 0.24]$; green,
                $\omega \in[2.4\times 10^{-3},2.4\times 10^{-2}]$;  blue, $\omega
                \in [2.4\times 10^{-4}, 2.4\times 10^{-3}]$; grey, $\omega \in
                [0,2.4\times 10^{-4}]$. Figure reprinted with permission from
                \cite{2006_prl_cheng}. Copyright (2006) American Physical Society.}\label{fig:intro:mri}
            \end{center}
        \end{figure}

        \noindent \textbf{Deep Layers -- }
        The symmetry breaking and the eventual disappearance of grain motion at the surface indicates
        that qualitatively different bulk flow is developing: the trumpet
        shape of the shear zones present in shallow layers must have
        changed. When $H/R_s$ is sufficiently large, the shear zone is
        entirely confined to the bulk of the material, and forms a
        dome-like structure above the rotating disk
        \cite{2004_prl_unger,2006_prl_fenistein,2006_prl_cheng,2008_epl_sakaie}
        -- see Fig.~\ref{fig:intro:mri}. The
        torsional failure of the material is thus concentrated in the
        dome. Cheng \textit{et. al.} measured the functional form of the
        axial velocity profile $\omega(z) |_{r=0}$, and found it to be
        described by a Gaussian:
        \begin{equation}
            \omega(z,r=0)=\omega_p+(1-\omega_p)\exp{-z^2/(2\sigma^2)}~,\label{eq:intro:axialslip}
        \end{equation}
        where $\omega_p$ is the rotation rate observed at the surface at
        $r=0$, which decreases roughly exponentially with $H$, and
        $\sigma$, the width of the bulk Gaussian velocity profile, is
        approximately $R_s/5$ \cite{2006_prl_cheng}. More recent experiments find
        a slightly different functional dependence on $z$ \cite{2009_subm_nichol}.\\

        \textbf{Couette versus split-bottom geometries -- }

\begin{figure}[tb!]
            \begin{center}
                \includegraphics[width=8cm]{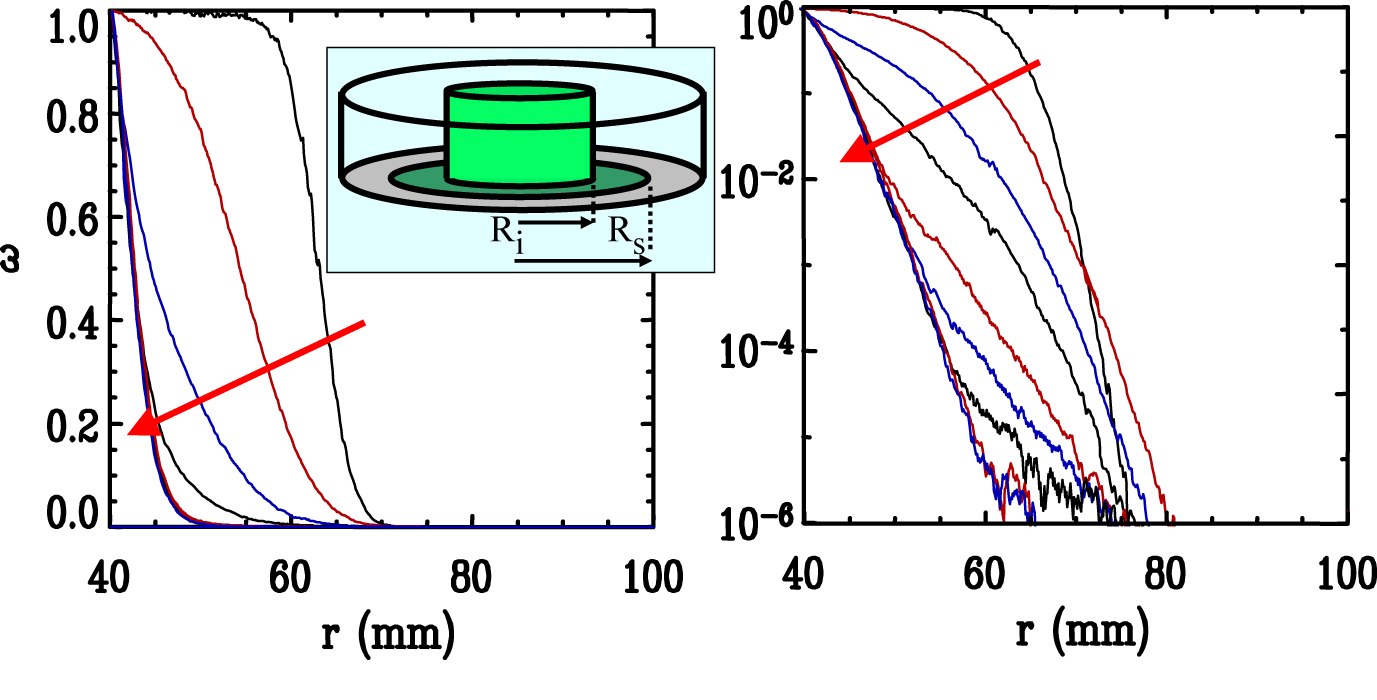}
                \caption{Surface flow
                profiles observed for 1 mm glass beads in a split-bottom Couette
                geometry, with inner cylinder radius $R_i=40$ mm, a split at $R_s=65$
                mm, and $ H=10,30,40,50,60,70,80,100,110$
                mm. The outer cylinder is 120 mm. Figure from Ref.~\cite{2005_thesis_mikkelsen}.}\label{fig:intro:couettesplibo}
            \end{center}
        \end{figure}

        The first studies~\cite{2003_nature_fenistein} of
        split-bottom geometries where done in a modified Couette cell, as
        shown in Fig.~\ref{fig:intro:couettesplibo}. The resulting flows are
        similar to the disc geometry, as long as the shear flow is away
        from the side walls, but since for increasing filling heights the
        shear zones move inward, they will inevitably 'collide' with the
        inner cylinder for sufficiently large filling height. The
        resulting flow profiles are shown in Fig.~\ref{fig:intro:couettesplibo}
        \cite{2005_thesis_mikkelsen}.

        First, one observes that for sufficiently large $H$, the flow
        profiles become independent of $H$. The main result is that the
        tail of these flow profiles becomes purely exponential for large
        $H$, while it is Gaussian for shallow $H$. We have found this
        exponential tail to be robust, i.e., independent of grain shape.
        Note that this does not contradict the findings of Mueth {\em et
        al.} -- these concern the shape of the flow profile near the
        shearing wall, corresponding to the range $10^{-3}< \omega < 1$
        -- flow profiles are indeed grain dependent here. But further out
        in the tail they become purely exponential. For other examples of
        exponential tails see
        \cite{2008_jstatmech_crassous}.\\

        \subsection{Dilatancy}\label{subsec:intro:dilate}

        By means of MRI, direct measurements
        of the evolution of the local packing density
        of the shear flow generated in a cylindrical split-bottomed
        geometry were performed in \cite{2008_epl_sakaie}. To be able to image the
        particles by means of MRI, food grade poppy seeds were used; these contain MRI-detectable
        mineral oils.

        It was observed that the relative change in density in the flowing
        zone is rather strong and saturates around 10-15 \%
        -- likely the rough and peanut shape of the poppy seeds plays a
        role. After long times, a large zone with essentially constant
        low packing fraction forms, encompassing most of the shearband. The
        fact that the density remains constant here, even though local
        strain rates vary over many decades, suggests that the density of
        the flowing material is a function of the \textit{total strain},
        and not of the strain rate \cite{2009_prl_kabla}.

        \begin{figure}[tb!]
            \begin{center}
                \includegraphics[width=8cm]{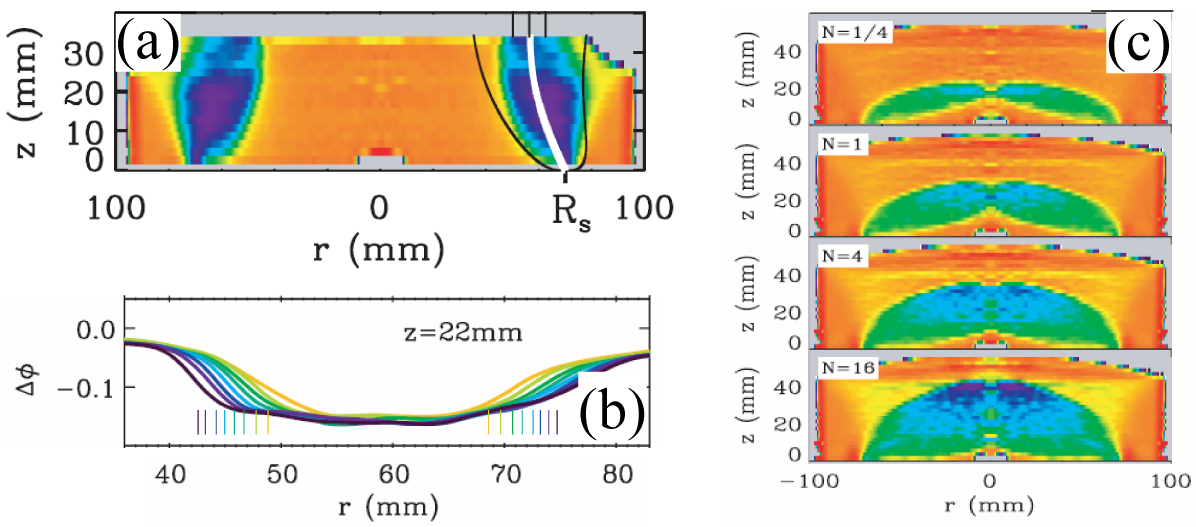}
                \caption{Evolution of
                dilatancy. (a) Color map of relative density change (light blue
                corresponds to -10\%) for $H/R_s =0.51$ after 4 rotations of the
                bottom disk (b) Spread of dilated zone for vertical shear observed
                in the bulk at $H/R_s =0.51$ at  $z=22$ mm ($H=36$ mm), for
                $N=1/2,1,2,\dots,64$, compared to estimates where, after $N$
                turns, the local strain equals one. (c) Spread of dilatancy for
                dome-like flow observed at $H/R_s =0.77$, for number of disc
                rotations $N$ as indicated. Figure adapted from Ref.~\cite{2008_epl_sakaie}. }
                \label{fig:intro:dilate}
            \end{center}
        \end{figure}

        Consistent with this, the dilated zone was found to slowly spread
        throughout the system as time progresses (Fig.~\ref{fig:intro:dilate}). This spread is
        consistent with the idea that, after initial preparation, the
        accumulated local strain governs the amount of dilatancy. To
        show this, the flow field in the bulk was reconstructed by
        combining the above mentioned scaling relations for the location
        and width of the shear zones in the bulk, and this reconstructed
        flow field can then be compared to the density field obtained by
        MRI. Such comparison shows that the locations of the dilated zone
        and the shear zone coincide, for small filling heights
        ($H/R_s<0.6$). Finally, for deep filling heights where torsional
        failure and precession play a role, a relatively long-lived
        transient was found to cause the dilated zone to deviate
        substantially from the late-time shear zone.

\section{Flows in the split-bottom geometry: Theory}\label{sec:intro:theory}

The flows in split-bottom geometries have been simulated both by
molecular dynamics simulations
\cite{2006_prl_cheng,2007_epl_depken,2008_partscitech_luding} as
well as by contact dynamics \cite{2007_pre_ries}, and a number of
theoretical approaches have been put forward. It remains
remarkable that no single consistent theoretical framework is
available from which Eqs.~\ref{eq:intro:erf}-\ref{eq:intro:weq}
can be deduced. Creating such a theory would constitute an
important milestone in the development of our understanding of
slow granular flows. Here we discuss the main approaches to flows
in the split-bottom geometry.\\

    \textbf{Variational principle -- } The first attempt to describe the flows in split-bottomed
    geometries goes back to Unger and
    coworkers \cite{2004_prl_unger}. The flows are treated in a
    Mohr-Coulomb fashion, with shear bands of zero width. The idea is
    to minimize the energy dissipation needed to sustain the flow.
    Calculating the total friction along the shear-sheet $r(z)$ using
    assumptions of constant friction coefficient $\mu$ and hydrostatic
    pressure $P$, amounts to finding the minimum of the functional
    (see Fig.~9)

    \begin{equation}
        T(H) = 2g\pi\rho\phi\mu\int_0^H (H-z)r^2 \sqrt{1+(dr/dz)^2}dz\label{eq:intro:minim}
    \end{equation}

    \noindent Here $\rho$ is the bulk density of the particles, $\phi$ is the
    average packing fraction ($\sim
    0.59$~\cite{2008_annurevfluid_forterre}) and $\mu$ is the
    effective friction coefficient. Identifying $r(z)$ with the center of the shearband $R_c(z)$,
    minimizing Eq.~\ref{eq:intro:minim} gives predictions for the location of the shearbands
    in the split-bottom geometry. The location of the shear zones predicted for shallow layers are
    very close to those observed, and for deep layers the model
    predicts a hysteretic transition to dome-like shear. Hence, while
    a number of aspects of split-bottom flows can be captured
    by this simple model, the hysteretic transition and zero width of
    the shear bands are clearly in contrast to experimental
    findings.\\

    \textbf{Extending the variational principle --}
    To explain the broad shearbands in the split-bottom geometry, a
    random  or randomly varying local material failure
    strength~\cite{2007_pre_torok, 2008_pre_jagla} was invoked. The
    main extension extends the minimal dissipation
    model by combining the variational principle with a
    self organized random potential as follows. At any given time, the
    shear band is represented as having zero width. However, the
    granular material is now taken to be inhomogeneous, with a local
    strength field which varies with space and is updated every
    timestep. At any given time the surface that minimizes the torque
    can be obtained, after which the strength field is updated etc. A
    smooth flow profile is then obtained by averaging over the
    different shear bands. The resulting flow fields are very close to
    those observed experimentally, with the only adjustable parameter
    controlling the effect of disorder. One possible point of
    criticism is that the model assumes that the fluctuating shear
    bands are radially symmetric -- see \cite{2008_pre_jagla} for
    details.

    \begin{figure}[tb!]
        \begin{center}
            \includegraphics[width=8cm, clip=true]{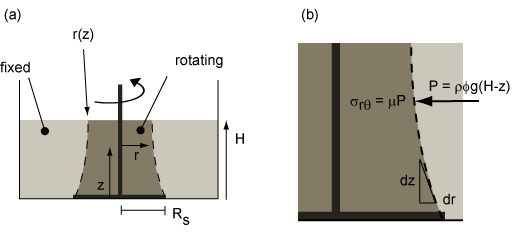}
            \caption{(a) An arbitrary shear zone of zero width $r(z)$ separates
            a rotating inner core and a static outer body. In (b): The frictional stress
            $\sigma_{r\theta} = \mu P$ on the shearing surface can be integrated to give the total
            driving torque necessary to rotate the disk in red. The integral
            is given in Eq.~\ref{eq:intro:minim}.}\label{fig:intro:ungerpic}
        \end{center}
    \end{figure}

    The main findings of \cite{2007_pre_torok} where confirmed by
    a different but related approach by
    Jagla~\cite{2008_pre_jagla}, and recently
    a two dimensional model using stochastic interparticle
    forces was shown to be able to also generate shear bands of finite
    width~\cite{2009_physa_bordignon}.\\

    \textbf{Inertial flows --} Jop performed simulations of flows in a
    cylindrical split-bottom setup
    ~\cite{2008_pre_jop}, using the
    inertial number theory, which should be valid for faster flows ~\cite{2008_annurevfluid_forterre}.
    The location of the shear zones in the
    bulk, the smooth transition to precession and the dome flows were
    all recovered. The width of the shear zones was found to scale
    with driving rate as $\Omega^{0.38}$, and indeed for slow flows
    the shear zones attain zero width. The inertial model therefore
    does not fully capture the physics of the slow split-bottom flows,
    but it does slightly better than Ungers original model~\cite{2004_prl_unger}
    in that it captures the smooth transition to precession.
    The only experimental work so far on rate dependent flows in the split-bottom is
    the work by Corwin~\cite{2008_pre_corwin}. He however described
    flows for which the centripetal forces dominate the
    gravitational and hydrostatic forces, which are outside the regime of
    flow rates studies by Jop.\\

    \begin{figure}[tb!]
        \begin{center}
            \includegraphics[width=8cm]{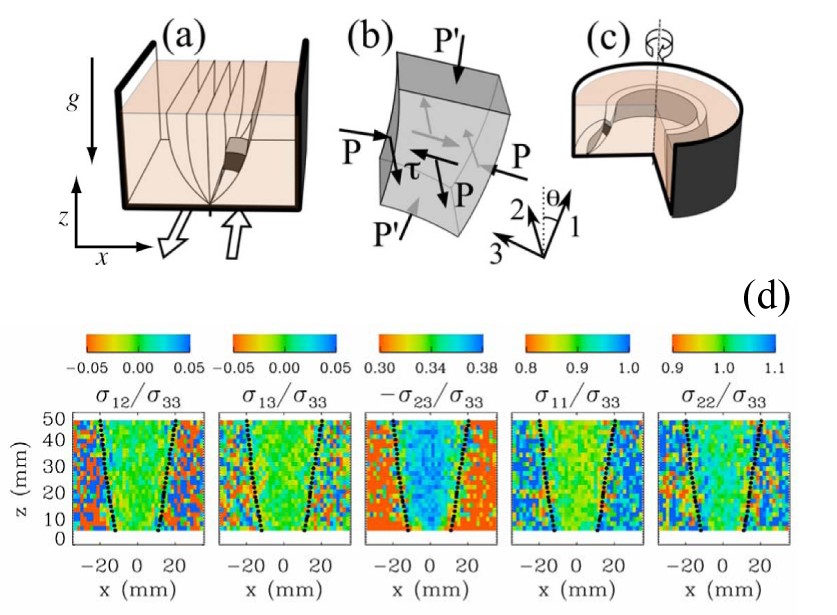}
            \caption{(a) Shear free sheets in a linear shear geometry. (b)
            Stress components acting in material (c) Shear free sheets in
            curved geometry. (d) Stress ratios in section of linear
            geometry, showing that the stresses are of the form of
            Eq.~\ref{eq:intro:depkenstress}, and that the friction coefficient
            $\mu=-\sigma_{23}/\sigma_{33}$ is not completely constant. Figure adapted from
            Ref.~\cite{2007_epl_depken}.}
            \label{fig:intro:dep}
        \end{center}
    \end{figure}

    \textbf{Shear free sheets -- }
    The form of the stress tensor in slow granular flows is a
    matter of debate. From a soil mechanics
    perspective, there is no reason to assume that the
    principle directions of stress and strain rate align, but Depken and
    coworkers have suggested that once there is flow in the system,
    the principle directions {\em do} align. The idea is that flow,
    even at a distance, creates sufficient amount of agitation that any
    amount of shear stress should lead to flow -- in other words,
    once there is flow, there is no clear yielding threshold anymore
    \cite{2009_subm_nichol,2006_pre_depken,2007_epl_depken}.

    Coaxiality severely restricts the form of the stress tensor, and
    for steady grain flows the flow can be decomposed into
    so-called shear free sheets  (SFS), that slide past each other --
    there is no (average) grain motion within the sheets
    (Fig.~\ref{fig:intro:dep}a-c). In this SFS basis, the stress tensor
    takes the form:

    \begin{equation}
        \sigma_{{\rm SFS}}=\left(\begin{array}{ccc}P'&0&0\\0&P&\tau\\0&\tau&P\end{array}\right)~.
        \label{eq:intro:depkenstress}
    \end{equation}

    The form of the stress tensor is reminiscent of that of the
    inertial model \cite{2006_nature_jop}, but with the exception that
    the stresses $P$ and $P'$ do not need to be equal (they are not in
    fact -- see Fig.~\ref{fig:intro:dep}), and that the friction coefficient
    $\mu:=\tau/P$ does not depend on flow rate, since the flows are
    rate-independent. However, as Depken showed \cite{2006_pre_depken},
    $\mu$ cannot be constant if the shear zones have finite width, and
    in fact has to attain a local maximum within the shear zone
    -- as subsequent numerical simulations indeed found
    \cite{2007_epl_depken,2007_pre_ries}.

    Recent contact dynamics simulations of Ries {\em et al.} in linear
    split-bottom cells \cite{2007_pre_ries} have confirmed that the stress
    and strain tensors align, so that the stresses take the form given
    by Eq.~\ref{eq:intro:depkenstress}. The
    alignment also occurs in the absence of gravity (to carry out
    these simulations, a mirror of the system was added; see Fig.~7 in ~\cite{2007_pre_ries}). Perhaps
    surprisingly, gravity appears not to be important for the
    understanding of split-bottom flows.\\

    \textbf{Spot model ---}
    To describe rate-independent flows in general, Bazant and coworkers
    put forward the 'spot'-model,
    which is based on the assumption that slow, dense
    granular flows are best described with a diffusion
    of low density regions in the material, called
    \emph{spots}~\cite{1972_japplphys_mullins}. The two-dimensional model
    describes flow profiles in chute flow and Couette geometries reasonably
    well.
    However, recently it was shown~\cite{2009_epje_woldhuis} that
    the very structure of this model is incapable of capturing the observed wide shearbands --
    the model's only lengthscale is the spot size, and
    therefore cannot yield wide shearbands. Moreover, its use of
    Mohr-Coulomb plasticity theory is in conflict with the
    observed co-axiality~\cite{2007_epl_depken,2007_pre_ries} of the stress and strain tensors.\\

\section{Outlook}

The surprising, robust and universal properties of slow granular
flows in split-bottom geometries have made the split-bottom
geometry into a versatile testing ground, not only for models of
slow flows, but also for experimental studies of related flows, of
mixtures and segregation ~\cite{2008_prl_hill}, of non-local
flows~\cite{2009_subm_nichol}, of faster grain flows
\cite{2008_pre_corwin,2009_thesis_dijksman} and of suspension
flows \cite{2010_inprep_dijksman}.

So far, no single convincing continuum theory to describe the wide
shear zones generated in split-bottom shear cells has been put
forward --- even though the experimental results strongly suggest
that these flows {\em should} be amenable to a continuum
description. The failure of the Mohr-Coulomb approach to describe
the internal structure of {\em narrow} shear bands might not be
troublesome, but its failure to describe these much wider shear
zones is cause for concern.

Apart from the theoretical challenges and experimental use of the
split-bottom geometry, the broad shear zones occurring in this
geometry allow for further experiments on slow flows, that are
more difficult to realize in narrowly localized shear bands. Open
questions for the future include to understand the microscopic
organization and velocity fluctuations in these flows, to
understand the role of interstitial fluid and grain shape, and to
explore the range of much smaller, but in particular much faster
driving rates.

\bibliographystyle{rsc}

\bibliography{softmatter_review}

\end{document}